\documentclass[twocolumn,pre,showpacs,superscriptaddress]{revtex4}
\pdfoutput=1 
\usepackage{graphicx}%
\usepackage{color} 
\usepackage{transparent}
\usepackage{psfrag}
\usepackage{dcolumn}
\usepackage{amsmath}
\usepackage{amssymb}
\usepackage{latexsym}
\usepackage{hyperref} 
\usepackage{booktabs}
\usepackage{latexsym}
\begin{document}

\def\bea{\begin{eqnarray}}
\def\eea{\end{eqnarray}}
\def\beq{\begin{equation}}
\def\eeq{\end{equation}}
\def\f{\frac}
\def\k{\kappa}
\def\e{\epsilon}
\def\be{\beta}
\def\D{\Delta}
\def\th{\theta}
\def\t{\tau}
\def\a{\alpha}
\def\J{{\cal J}}

\def\cDa{{\cal D}[X]}
\def\cDd{{\cal D}[X^\dagger]}
\def\cL{{\cal L}}
\def\cLo{{\cal L}_0}
\def\cLa{{\cal L}_1}

\def\Re{{\rm Re}}
\def\sj{\sum_{j=1}^2}
\def\rk{\rho^{ (k) }}
\def\rek{\rho^{ (1) }}
\def\cek{C^{ (1) }}
\def\rz{\rho^{ (0) }}
\def\rt{\rho^{ (2) }}
\def\rtb{\bar \rho^{ (2) }}
\def\trk{\tilde\rho^{ (k) }}
\def\trek{\tilde\rho^{ (1) }}
\def\trz{\tilde\rho^{ (0) }}
\def\trt{\tilde\rho^{ (2) }}
\def\r{\rho}
\def\tD{\tilde {D}}
\def\C{{\cal {C}}}

\def\s{\sigma}
\def\kb{k_B}
\def\F{{\cal F}}
\def\la{\langle}
\def\ra{\rangle}
\def\nn{\nonumber}
\def\up{\uparrow}
\def\dn{\downarrow}
\def\S{\Sigma}
\def\dg{\dagger}
\def\d{\delta}
\def\p{\partial}
\def\l{\lambda}
\def\le{\left}
\def\ri{\right}
\def\L{\Lambda}
\def\G{\Gamma}
\def\o{\Omega}
\def\w{\omega}
\def\g{\gamma}
\def\mt{\tilde {m} }
\def\mbt{\tilde {\bf m} }

\def\jv{ {\bf j}}
\def\jr{ {\bf j}_r}
\def\jd{ {\bf j}_d}
\def\noi{\noindent}
\def\a{\alpha}
\def\d{\delta}
\def\p{\partial} 

\def\H{ {\bf H}}
\def\He{{\bf H_e}}
\def\h{{\bf h}}
\def\m{{\bf m}}
\def\hth{h_{\theta}}

\def\la{\langle}
\def\ra{\rangle}
\def\e{\epsilon}
\def\n{\eta}
\def\g{\gamma}
\def\break#1{\pagebreak \vspace*{#1}}
\def\hf{\frac{1}{2}}

\title{Rotational diffusion under torque:  Microscopic reversibility and excess entropy} 
\author{Swarnali Bandopadhyay}
\email{swarnalib@tifrh.res.in}
\affiliation{TIFR Centre for Interdisciplinary Sciences, 
21 Brundavan Colony, Narsingi, Hyderabad 500075, Telengana, India
}
\author{Debasish Chaudhuri}
\email{debc@iopb.res.in}
\affiliation{Institute of Physics, Sachivalaya Marg, Bhubaneswar 751005, India
}
\author{A. M. Jayannavar}
\email{jayan@iopb.res.in}
\affiliation{Institute of Physics, Sachivalaya Marg, Bhubaneswar 751005, India
}
\date{\today}

\begin{abstract}
We consider rotational diffusion for two systems - a macrospin under external magnetic field, and a particle diffusing on the surface of a sphere under external torque. Microstates in the two cases transform differently under time-reversal. This results in Clausius like dependence of stochastic entropy production (EP) for macrospins, and an excess EP for diffusion of particles on sphere. The total EP in both the cases obey fluctuation theorems. For macrospins, we derive analytical expression for probability distribution of total EP in  the adiabatic limit.  
Numerical simulations show that the distribution functions of EP agree well with theoretical predictions. 
\end{abstract}


\maketitle
\section{Introduction}
Stochastic thermodynamics has extended the definitions of thermodynamic quantities like work, energy, entropy etc. to their stochastic counterpart, as a description of stochastic evolution of non-equilibrium systems with small degrees of freedom~\cite{Jarzynski2011,Seifert2012}. This allows one to obtain energy balance, and equalities involving entropy production (EP), or work done known as fluctuation theorems (FT)~\cite{Evans1993, Gallavotti1995, Jarzynski1997, Sekimoto1998, Lebowitz1999, Crooks1999, Seifert2005, Baiesi2009, Baiesi2010, Hummer2010, Kurchan2007, Narayan2004, Jayannavar2007, Lahiri2014a, Saha2009, Lahiri2009,Sahoo2011}. In experiments FT symmetries were observed~\cite{Wang2002,Blickle2006,Speck2007,Joubaud2012}, and used to extract free energies from non-equilibrium measurements~\cite{Liphardt2002,Collin2005}. The ideas of stochastic thermodynamics have been extended to active particles as well~\cite{Hayashi2010,Seifert2011,Ganguly2013,Chaudhuri2014}.

Stochastic energy balance can be derived from appropriate Langevin equations giving the definition of dissipated heat. This does not depend on how the microscopic dynamical variables transform under time-reversal operation. However, EP captures breaking of time-reversal symmetry, and does depend on how microstates transform under time-reversal operation. In this paper, using two systems whose dynamics are given by the same equation of motion, but whose microstates transform differently under time-reversal, we demonstrate how the EP in them are different. While one of these systems show stochastic reservoir EP consistent with Clausius expression, the other gives rise to an excess EP which can not be captured by the dissipated heat. We focus on stochastic thermodynamics of macrospins having a single magnetic domain, and a related system of particles diffusing on the surface of a unit sphere.  

With advent of spintronics, magnetic devices are getting miniaturized.  
In such devices, macrospins reside in a complex magnetic environment that may produce time varying torque. 
The impact of thermal fluctuations increases inversely with reducing size of devices~\cite{Blanter2000,Jr1963,Coffey2012}, giving rise to drastic effects like magnetization reversal~\cite{Parkin2000}.  Several recent studies focussed on how to control magnetic devices against thermal noise~\cite{Tserkovnyak2001,Foros2005,Foros2007,Bandopadhyay2011,Covington,Utsumi2015}. 
In Ising spins following Glauber dynamics, distribution of dissipative work was presented in Ref.~\cite{Marathe2005}.
Unlike Ising spins, macrospins in presence of magnetic fields undergo stochastic rotational motion.
The diffusion of a particle on unit sphere in presence of external torque, is described by Langevin equations closely related to that describing the stochastic macrospin dynamics.
However, the origin of torque does not anymore come from a conservative magnetic energy density, rather is imposed externally. Also, the notion of dissipative and reactive currents depend on the transformation of angular positions identifying microstates. While magnetic field and magnetization are odd variables under time reversal, angular position of diffusing particle and external torque are even variables, leading to different forms of EP.

\section{Macrospin}
First let us consider a macrospin with magnetization $\m$ in presence of a time-dependent magnetic field $\H(t)$. The deterministic dynamics  $\dot \m = \g \m \times \H(t)$, where $\dot\m = d\m/dt$, and $\g$ denotes the gyromagnetic ratio, conserves magnetization $d m^2/dt=\m.\dot\m =0$. For a time-independent field $\H$, the macrospin precesses around the field due to spin torque  $\m \times\H$. 
Stochastic dynamics of macrospin may involve fluctuations in both amplitude and direction of magnetization~\cite{Ma2012,Bandopadhyay2015}. However, for materials with high enough Curie temperatures, one may neglect the amplitude fluctuation~\cite{Jr1963,Jayannavar1991,Seshadri1982, Bandopadhyay2015a}. This naturally leads to a Langevin dynamics known as the Landau-Lifshitz-Gilbert equation which involves a multiplicative noise. 
The macrospin coupled to a heat bath gets influence from the heat bath in terms of
a stochastic field $\h(t)$ and a related dissipation with a damping coefficient $\eta$ such that~\cite{Kubo1962,Jr1963,Kubo1970}
\begin{equation}
\dot \m=\gamma \, \m \,\times\, \left[\H+\h(t)-\eta\dot \m \right].
\label{LLG1}
\end{equation}
The stochastic magnetic field obeys Gaussian statistics with 
 \bea
 \la \h (t) \ra  &=& 0, \nn\\ 
 \la \h (t) \otimes \h(t')\ra &=& 2  D_0 {\bf 1} \d(t-t')
 \label{noise}
 \eea
 where ${\bf 1}$ denotes the identity matrix, $D_0=\eta \kb T/V$ with $T$ denoting the temperature, $\kb$ the Boltzmann constant, and $V$ the volume of the magnetic particle. The LLG equation may be derived using the Zwanzig formalism, coupling the macrospin with a heat bath composed of either spins~\cite{Seshadri1982} or harmonic oscillators~\cite{Jayannavar1991}. The magnetic field $\H$ is obtainable from an energy density $G = -\m.\H$ by using $\H=-\p G/\p \m$. 

The rotational diffusion of the orientation $\m$ on the surface of a sphere of radius $m$ may be represented in terms of angular position $[\theta (t),\phi (t)]$. The Langevin dynamics can then be expressed as 
\bea
\dot\th &=& h'\, m( H_\th+ h_\th) - g' m (\sin\th)^{-1} (H'_\phi +  h_\phi) \nn\\
\sin\th \, \dot \phi &=& g'  m(H_\th +   h_\th) + h' m (\sin\th )^{-1}  (H'_\phi +  h_\phi),\nn\\
\label{llg2}
\eea
where
\bea 
 g' = \f{1/\g m}{ (1/\g^2)+\eta^2 m^2 },\, h'=\f{\eta}{(1/\g^2)+\eta^2 m^2}. \nn
\eea
In Eq.(\ref{llg2}), $\dot\th=\p_t \th$, $\dot\phi=\p_t\phi$, and $\H = \hat \th H_\th + \hat \phi H_\phi$  with
$H_\th = -({1}/{m})\p_\th G$, $H'_\phi \equiv H_\phi \sin\th = -({1}/{m})\p_\phi G$. The angular components of stochastic field can be expressed in terms of their cartesian components as
$h_\th = h_x \cos\th \cos\phi+ h_y \cos\th \sin\phi - h_z\sin\th$, 
and $h_\phi = -h_x\sin\th \sin\phi + h_y \sin\th \cos\phi$. 
Note that Eq.(\ref{llg2}) involves multiplicative noise. Recently Ref.~\cite{Aron2014} showed explicitly that the form of FP equation derived from the LLG equation is independent of the choice of stochastic calculus -- Ito, Stratonovich or a post-point discretization scheme~\cite{Lau2007a, vanKampen1992}. In the following, we use this FP equation which was originally derived in~\cite{Jr1963} using the Stratonovich convention that we use  throughout this paper.

The FP equation corresponding to Eq.(\ref{llg2}) has the form
\bea
\p_t P = -{\bf \nabla}_{\o}. {\bf J}_\o,~~  {\bf J}_\o =\hat\th J_\th + \hat \phi J_\phi  
\label{fpeq}
\eea
where the divergence of current on the right hand side is given by
${\bf \nabla}_{\o} . {\bf J}_\o = \f{1}{\sin \th} \p_\th (\sin \th J_\th ) + \f{1}{\sin \th} \p_\phi J_\phi$, with $\o$ denoting the
solid angle. 
The two components of probability current are given by~\cite{Jr1963}
\bea
J_\th &=&  m [h' H_\th - g' H_\phi] P - k' \p_\th P \nn\\    
 J_\phi &=&   m [ g' H_\th + h' H_\phi] P - k' (\sin\th)^{-1} \p_\phi P. 
 \label{current}
\eea
Here $h'$ and $g'$ play the role of mobility, and $k'$ plays the role of diffusivity.  
These mobility and diffusivity coefficients obey Einstein relation 
$k'=D_0 m^2 (h'^2 + g'^2)= (\kb T/V) \, [\eta/(1/\g^2 +\eta^2 m^2)] = \kb T h'/V$~\cite{Jr1963}. 

Note that equations (\ref{llg2}) and  (\ref{fpeq}) also describe the motion of a particle diffusing on the surface of a sphere under position dependent external torque, with reinterpretation of some of the terms --
$m$ should be interpreted as the radius of the sphere,  $h'$ and $g'$ will mean mobility. In absence of $(H_\th, H_\phi)$ the  equations describe simple diffusion on a sphere of radius $m$. $[ H_\th, H_\phi]$ acts as an external torque which in general could be any function of $(\th,\phi,t)$, and need not be derivable from an energy density like $G$. For macrospins, $V$ denotes the total volume of the spin, which can be set to unity for particle diffusion, without any loss of generality. We discuss this dynamics in Sec.~\ref{sph_diffusion}.
Note that when treating Eq.s (\ref{llg2}) and  (\ref{fpeq})  as a description of magnetization dynamics, $\m$ and $\H$ has to be treated as odd parity variables under time reversal. This reflects in the way ($\th,\phi$) and $(H_\th, H_\phi)$ transform under time reversal. On the other hand, for a particle diffusing on the surface of a sphere,  position ($\th,\phi$)  are even parity variables under time reversal, and in that case $(H_\th, H_\phi)$ having the meaning of externally imposed torque does not change sign under time-reversal. This difference gives rise to two different expressions for EP in the two cases. While for macrospin dynamics one obtains Clausius like relation for EP in the reservoir, for rotation diffusion of particles one finds an {\em excess} EP apart from the Clausius term. We show this in detail in the following.

The  non-equilibrium Gibbs entropy is given by~\cite{Crooks1999, Seifert2005} 
$$S = -\kb \int d\o\, P(\th,\phi,t) \ln P(\th,\phi,t) = \la -\kb \ln P \ra,$$
where $\int d \o =\int \sin\th\, d\th\, d\phi\,$ is the integration over 
all possible solid angles, and $\la \dots \ra$ denotes statistical average. The above definition of $S$ is the same
as the Shanon information entropy of a given probability distribution~\cite{Shanon1948, Kardar2007}. 
The Szilard engine and Maxwell's daemon paradox~\cite{Szilard1929} helped building the connection between 
Shanon's information entropy and thermodynamic entropy~\cite{Maruyama2009,Mandal2012,Leff1990}. Note that Landauer's principle linked erasure of one bit of information with minimal heat dissipation by an amount $\kb T \ln 2$~\cite{Landauer1961}, and this has been experimentally verified~\cite{Berut2012}. The definition of entropy $S$ thus has a much wider scope, including a description of non-equilibrium processes. The stochastic entropy of a micro-state is given by $s(\th,\phi,t) = -\kb \ln P(\th,\phi,t)$, such that  $S = \la s \ra$.
One can express the rate of change in stochastic entropy as 
\bea
\f{\dot s}{\kb} &=& -\f{\p_t P}{P} - \f{\p_\th P}{P} \dot \th - \f{\p_\phi P}{P} \dot \phi. 
\label{sdot1}
\eea

We now consider the two cases of macrospin dynamics under external magnetic field, and diffusion of particle on a sphere in presence of torque, separately.

\subsection{Stochastic energy balance}
The rate of stochastic energy gain per unit volume $\dot G  = -\H \cdot \dot\m - \m \cdot \dot\H$,
the rate of work done $\dot W= -\m \cdot \dot\H$, stochastic heat absorption by the
system $\dot q = -\H \cdot \dot\m$. Thus the stochastic energy balance $\dot G = \dot q + \dot W$. 
In the spherical polar coordinate, 
\bea
\dot q = -\H\cdot \dot\m  &=& - [\hat\th H_\th  + \hat\phi H_\phi ] \cdot [\hat\th\, m \dot\th + \hat\phi\, m \sin\th \dot\phi]  \nn\\
&=&  - m \le[ H_\th \dot\th  + H_\phi \sin\th \,\dot\phi \ri] \, .
\label{heat_1}
\eea
Note that the rate of total heat absorption is given by $\dot Q = V \dot q$. 

\subsection{Entropy production using Fokker-Planck equation}
\label{ep_fp}

At this stage, it is crucial to identify the properties of microstate $\m$ and the probability current $ {\bf J}_\o$ under time reversal. As noted before, $\m(t)$ and $\H(t)$ are odd variables under time reversal. The $\m \to -\m$ operation is equivalent to taking the spatial configuration 
$(\th,\phi) \to (\pi-\th, \pi+\phi)$.  
These transformations lead to: $\sin\th \to \sin\th$, $\p_\th \to -\p_\th$ and $\p_\phi \to \p_\phi$; and as a result $H_\th \to -H_\th$ and $H_\phi \to H_\phi$. 

The original FP equation can be expressed as $\p_t P = -{\bf \nabla}_{\o}. ({\bf J}_\o^{(r)}+{\bf J}_\o^{(d)})$, where ${\bf J}_\o^{(r)}$ denotes reactive current and ${\bf J}_\o^{(d)}$ denotes dissipative current. Under time reversal one obtains $\p_t P = -{\bf \nabla}_{\o}. {\bf J}_\o^{(r)} +{\bf \nabla}_{\o}. {\bf J}_\o^{(d)}$, where ${\bf J}_\o^{(r)} = (-m g' H_\phi P, m g' H_\th P)$ and the
dissipative components of current:
\bea
J_\th^{(d)} &=&  m h' H_\th P - k' \p_\th P \nn\\    
 J_\phi^{(d)} &=&   m h' H_\phi P - k' (\sin\th)^{-1} \p_\phi P.
 \label{Jd}
\eea
Using Eq.(\ref{sdot1}) and expressing $\p_\th P$ and $\p_\phi P$ in terms of the dissipative currents one gets
\bea
\f{\dot s}{\kb} 
= -\f{\p_t P}{P} + \f{J_\th^{(d)} \dot \th + J_\phi^{(d)} \sin\th\, \dot\phi}{k' P} + \f{\dot Q}{\kb T}. 
\label{sdot}
\eea
In obtaining the third term on the right hand side of the above relation, we used  the identity $h'/k' = V/\kb T$.

At this point, we perform a two step averaging : (i)~over trajectories and (ii)~over 
the ensemble of all possible solid angles $\o$ with probability $P(\o,t)$. The trajectory average of the components of angular velocity leads to
$\la \dot \th |\, \th, \phi, t\ra = J_\th/P$ and   $\la \sin\th\, \dot \phi |\, \th, \phi, t\ra = J_\phi/P$~\cite{Seifert2005}. Note that $J_\th = J_\th^{(d)} - m g' H_\phi P$ and
$J_\phi = J_\phi^{(d)} + m g' H_\th P $. In order to perform averaging over the microstate probability $P(\o,t)$, we multiply Eq.(\ref{sdot}) throughout by $P(\o,t)$ and integrate over $\o$.
The conservation of probability $\int d\o P(\o,t)=1$ gives $\int d\o \p_t P(\o,t) = 0$. The resultant expression for the average EP in the system
\begin{align}
\f{\la \dot s\ra}{\kb} 
&=  \int d\o  \f{ (J_\th^{(d)})^2 + (J_\phi^{(d)})^2  }{k' P}  \nn \\ 
&+\f{m g'}{k'} \int d\o \le[ J_\phi^{(d)}H_\th -  J_\th^{(d)} H_\phi\ri] 
 + \f{\la \dot Q \ra}{\kb T}. 
\end{align}
Note that $\int d\o H_\th H_\phi P =0$, due to the inversion symmetry of $H_\th $, $H_\phi$ through the centre of the coordinate system. 
Also one can show that $\int d\o H_\phi \p_\th P =0$ and $\int d\o H_\th \p_\phi P =0$, using integration by parts. Thus the second term in the above equation vanishes, giving us
\bea
\dot S \equiv \la \dot s\ra &=& \kb  \int d\o  \f{ (J_\th^{(d)})^2 + (J_\phi^{(d)})^2  }{k' P} + \f{\la \dot Q \ra}{\kb T} \\
&=& \dot S_t - \dot S_r. \nn
\eea
Note that $\dot S_r = -\la \dot Q \ra/T$ is the entropy flux to the environment obeying Clausius theorem. The total average EP in system and environment is
\bea
\dot S_t = \dot S + \dot S_r =  \kb  \int d\o  \f{ (J_\th^{(d)})^2 + (J_\phi^{(d)})^2  }{k' P}  \geq 0
\label{dot_St}
\eea
in accordance with the second law of thermodynamics. 
Non-zero dissipative currents $J_\th^{(d)}$ and $J_\phi^{(d)}$ quantify the irreversible non-equilibrium processes taking place within the system. 
Calculations of thermodynamic EP using the FP equation have been presented in other contexts in Ref.~\cite{Tome2006,Tome2010,Tome2015}. 
The definition of stochastic entropy of the system $s = -\kb \ln P$, along with the FP equation gave us the stochastic reservoir EP 
\bea
\dot s_r = - \f{\dot Q}{T} = \f{mV}{T} \le[ H_\th \dot\th + H_\phi\, \sin\th \dot\phi \ri].
\label{srdot}
\eea  

\subsection{Detailed balance}
\label{eqm}
At equilibrium, due to time-reversibility, all the components of dissipative current has to vanish separately, such that the average total EP is zero. 
Considering a time-independent magnetic field applied along the $z$-axis, $G=-mH\cos\th$,
$H_\phi=0$. Then $J_\phi^{(d)} =0$ implies $P$ is independent of $\phi$. The other constraint $J_\th^{(d)} =0$ gives $dP/P=m (h'/k') H_\th d\th$. 
Integrating this equation, and Using the identities $H_\th = -H \sin\th$, $h'/k' = V/\kb T$, one obtains $P=\f{1}{Z(H)}\exp[- GV/\kb T ]$, where $Z(H)$ denotes the partition function at a given field strength $H$.

\subsection{Fluctuation theorems}
\label{ft_1}
Assume a macrospin evolves from $t=0$ to $\t_0$ along a trajectory $X = [\m(t), \H(t)]$ where time dependent field $\H(t)$ denotes the protocol of forward process. 
Let us divide the path into $i=1,2,\dots,N$ segments of time-interval $\d t$ with $N \d t = \t_0$.
The transition probability $p_i^+ (\th', \phi', t+\d t | \th, \phi,t)$ on $i$-th segment is controlled by the Gaussian random process with probability
$P({\bf h}_i) = (\d t/4\pi D_0)^{1/2} \exp(-\d t\, {\bf h}_i^2/4 D_0)$ where  ${\bf h}_i$ denotes the stochastic noise at $i$-th instant. 
Denoting Eq.(\ref{llg2}) as  
$\dot \th = \Theta(\th, \phi, H_\th, H_\phi)$ and $\dot \phi = \Phi(\th,\phi, H_\th,H_\phi)$, %
the transition probability on $i$-th segment
$p_i^+ = \J_i \la \d(\dot \th_i -\Theta_i) \d(\dot \phi_i - \Phi_i) \ra =  \J_i  \int d {\bf h}_i P({\bf h}_i) \d(\dot \th_i -\Theta_i) \d(\dot \phi_i - \Phi_i)$. 
The  Jacobian of transformation $\J_i={\rm det}[\p(h_{x_i}, h_{y_i}, h_{z_i})/\p(m_i, \th_i, \phi_i) ]_{m_i = {\rm constant}}$.
The probability of the complete path is ${\cal P}_+ = \prod_{i=1}^N p_i^+$.

\begin{figure}[t] 
\begin{center}
\includegraphics[width=8cm]{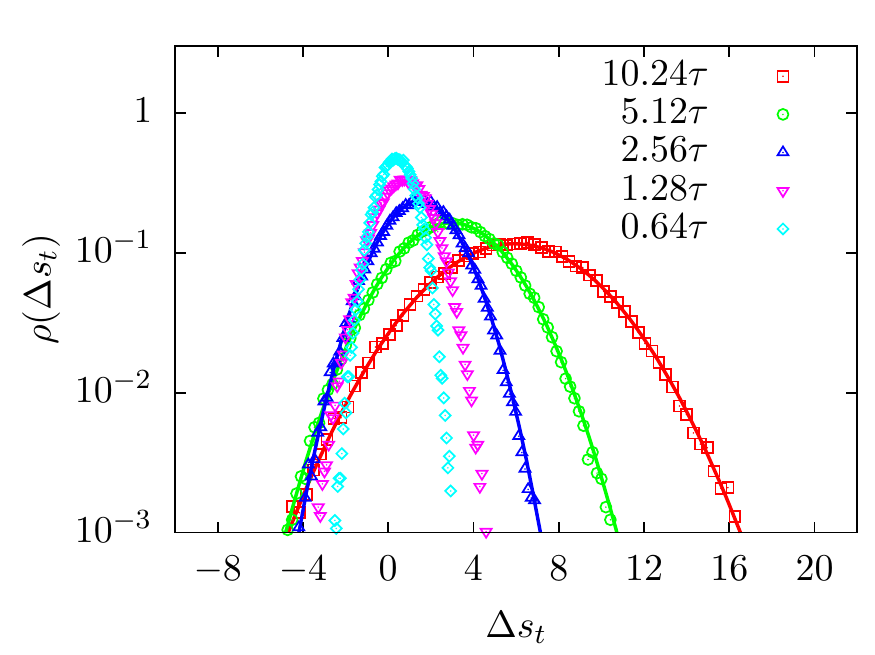}
\caption{(Color online) Distribution of entropy production for a macrospin initially in absence of magnetic field, driven by a linearly increasing field $H(t) =\a\, t$ with time, where the rate $\a = H_f/\t_0$ with final field strength $H_f = 0.1\, \kb T/m$  at $t=\t_0$. Different curves show entropy distributions calculated over various $\t_0$ denoted in the legend. For $\t_0 = 2.56, 5.12, 10.24\, \t$ 
the data fit with the analytic form given by Eq.(\ref{rho-gauss}) giving $\la \D s_t \ra=5.92, 2.96, 1.47\, \kb$ respectively.
}
\label{rho_st}
\end{center}
\end{figure}

As $\m$ and $\H$ are odd variables under time reversal, the time-reversed trajectory can be denoted as $X^\dagger = [-\m(\t_0-t), -\H(\t_0 -t)]$.  
Replacing $\th \to \pi-\th$, and $\phi \to \pi+\phi$ in Eq.(\ref{llg2}) one obtains the equation governing angular dynamics along time reversed trajectories. 
Denoting these equations as $\dot \th = \Theta^\dagger(\th, \phi, H_\th, H_\phi)$ and $\dot \phi = \Phi^\dagger(\th,\phi, H_\th,H_\phi)$,
the probability of conjugate trajectory  can be expressed as 
${\cal P}_- = \prod_{i=1}^N p_i^-$, where 
$p_i^- =    \J_i^- \la \d(\dot \th_i -\Theta^\dagger_i(\t_0 -t)\,)\, \d(\dot \phi_i - \Phi^\dagger_i(\t_0 -t) \ra $, 
where $ \J_i^-$ denotes the relevant Jacobian.  As $\J_i^-=\J_i$, Jacobians drop out of the ratio $p_i^+ / p_i^-$~\cite{Bandopadhyay2015}. 
Thus one obtains
\bea
\f{\D s_r}{\kb} = \ln\f{{\cal P}_+}{{\cal P}_-} = \f{V\,m}{\kb T} \int_0^{\t_0} dt  \le[ H_\th \dot\th + H_\phi \sin\th \dot\phi \ri].
\label{dsr}                                        
\eea
The above expression of $\D s_r$ corresponds to $\dot s_r$ derived from FP equation [see Eq.(\ref{srdot})].
The trajectories considered above describe evolution from a distribution of initial states $P_i$ to that of final states  $P_\ell$, with the change in 
system entropy 
\beq
\D s 
= \kb \ln (P_i/P_\ell).
\label{ds}
\eeq 
Thus the total entropy change $\D s_t  =  \D s + \D s_r = \kb \ln  \f{{\cal P}^f [X] }{{\cal P}^b [X^\dagger]} $, 
where total probabilities of time forward and time reversed trajectories are given by ${\cal P}^f [X] = P_i {\cal P}_+$ and ${\cal P}^b [X^\dagger] = P_\ell {\cal P}_-$ respectively.
As the Jacobian of transformation from $X$ to $X^\dagger$ is unity, one readily gets the integral fluctuation theorem (IFT) $\la e^{-\D s_t/\kb} \ra = 1$, leading to $\la \D s_t \ra \geq 0$
via Jensen inequality.

In a steady state the total entropy change $\D s_t$ along a time-forward path $\D s^f_t(X)$ is equal and opposite to that along the time-reversed path, $\D s^b_t(X^\dagger) = -\D s^f_t(X)$. 
This leads to the detailed fluctuation theorem (DFT)~\cite{Kurchan2007,Crooks1999} 
\bea
\r(\D s_t) &=& e^{\D s_t/\kb} \rho(-\D s_t).
\eea

{
\subsection{Distribution of entropy production}
\label{simu}
Let us consider a time dependent magnetic field $\H = H(t) \hat z$ with linear time-dependence $H(t)=H_0 +\a t$. Writing it in a dimensionless form $m H/\kb T = \e (1+\zeta t/\t)$ where $\e = mH_0/\kb T$,
and $\zeta = \a \t/H_0$ is the dimensionless rate of change of the field, with unit of time set by $\t=m/\g \kb T$. 
For slow rate $\zeta \ll 1$ the system remains close to equilibrium, and for fast rate $\zeta \gg 1$ the variation in magnetic field is too fast for the instantaneous magnetization to follow it. 
From numerical simulations using Stratonovich discretization of  Eq.(\ref{llg2}) with time step $\d t = 0.01 \t$ we calculate fluctuations in EP, and  
obtain its probability distributions at various driving rates. 
We express the magnetization in units of $m$,  and energy in units of  $\kb T$. In calculating total EP we use the expressions given in Eq.s  (\ref{dsr}) and (\ref{ds}).

\begin{figure}[t] 
\begin{center}
\includegraphics[width=8cm]{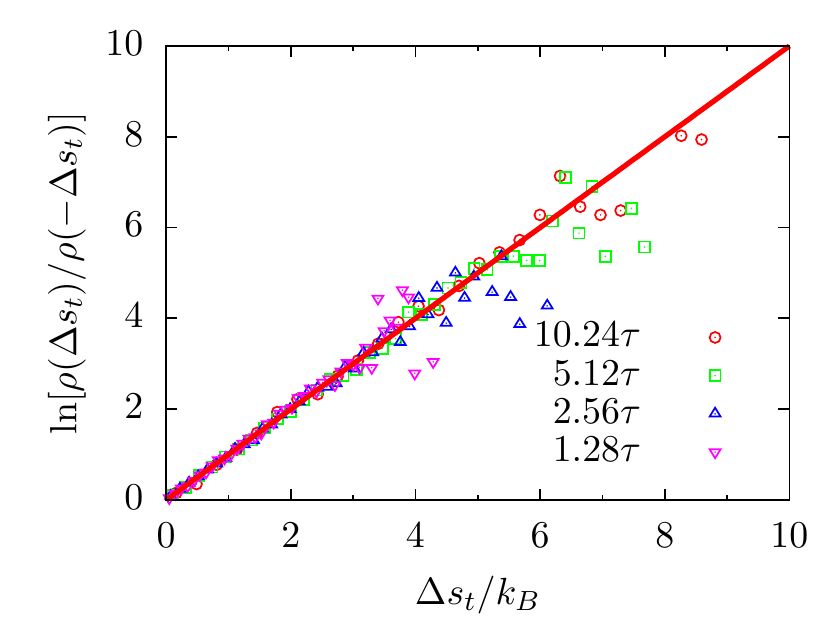}
\caption{(Color online) Ratio of probability distributions of positive and negative EP, $\ln [\r(\D s_t)/ \r(-\D s_t)]$ calculated from the data 
shown in Fig.~\ref{rho_st}. The solid line is a plot of the function $\D s_t/\kb$. The deviation of data for larger $\D s_t$ is due to lack of statistics.
}
\label{PSratio}
\end{center}
\end{figure}

Let us change the magnetic field from $0$ to $H_f=0.1$ (in units of $\kb T/m$) in a time window $\t_0$, which sets the rate $\a=H_f/\t_0$.  
The distribution functions for different rates are plotted in Fig.\ref{rho_st}, and are denoted by the values of $\t_0$. In Fig.~\ref{PSratio}, the natural logarithm of the ratio of probabilities of 
positive and negative  EP is plotted against EP. This shows good agreement with the prediction of detailed FT.  
Note that, for slower driving rates ($\t_0 \geq 2.56\, \t$), the distribution functions are broad, and have Gaussian profile (Fig.\ref{rho_st}). The Gaussian nature can be understood by splitting the total time $\t_0$ into smaller intervals, beyond which fluctuations of magnetization are not correlated. Since the reservoir EP is a sum over many such uncorrelated random events [Eq.(\ref{dsr}],  one obtains Gaussian distribution in accordance with the central limit theorem, 
\beq
\rho(\D s_t) = \f{1}{\sqrt{2\pi \s^2}} e^{-(\D s_t - \la \D s_t \ra)^2/2\s^2}, \nn
\eeq 
peaked at the mean EP $\la \D s_t \ra$. Moreover, the distribution should obey the IFT, $\la e^{-\D s_t/\kb} \ra =1$. 
This requires $\s^2 = 2 \kb \la \D s_t \ra$. Thus the distribution of EP is expected to have the form
\beq
\rho(\D s_t) = \f{1}{\sqrt{4\pi \kb  \la \D s_t \ra}} e^{-(\D s_t - \la \D s_t \ra)^2/4 \kb \la \D s_t \ra}.
\label{rho-gauss}
\eeq 
The lines in Fig.\ref{rho_st} show fit of numerical data to this function.

\begin{figure}[t] 
\begin{center}
\includegraphics[width=8cm]{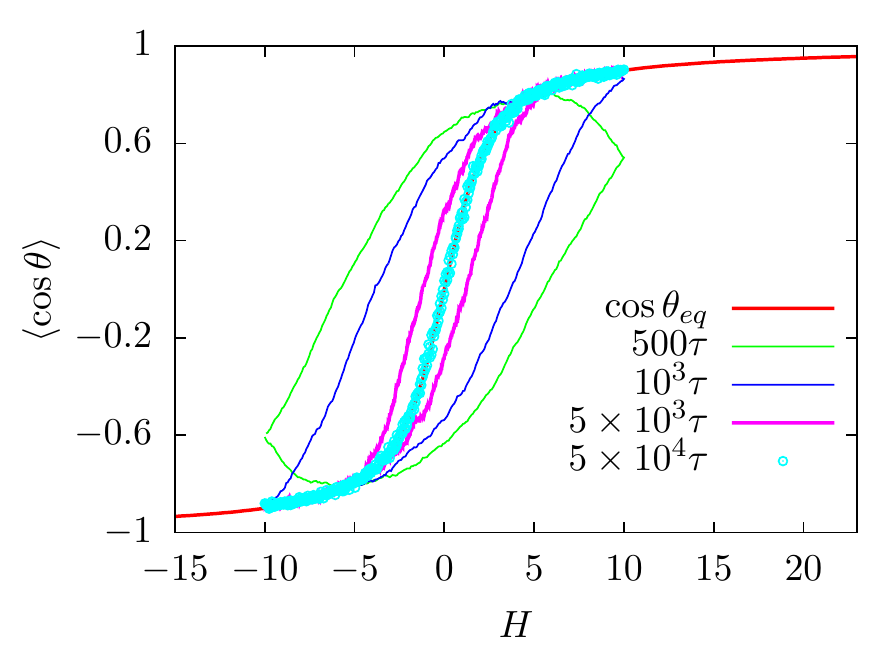}
\caption{(Color online) Hysteresis curves at non-equilibrium steady states. The curve $\cos \th_{\rm eq}$ plots the equilibrium Langevin function.
The legend denotes time period of changing external field. As the cycle gets extremely slow ($t_p=5\times 10^4\t$), $m_z/m=\la \cos\th\ra$ 
collapses onto the equilibrium curve.  
}
\label{hysteresis}
\end{center}
\end{figure}

\begin{figure}[t] 
\begin{center}
\includegraphics[width=8cm]{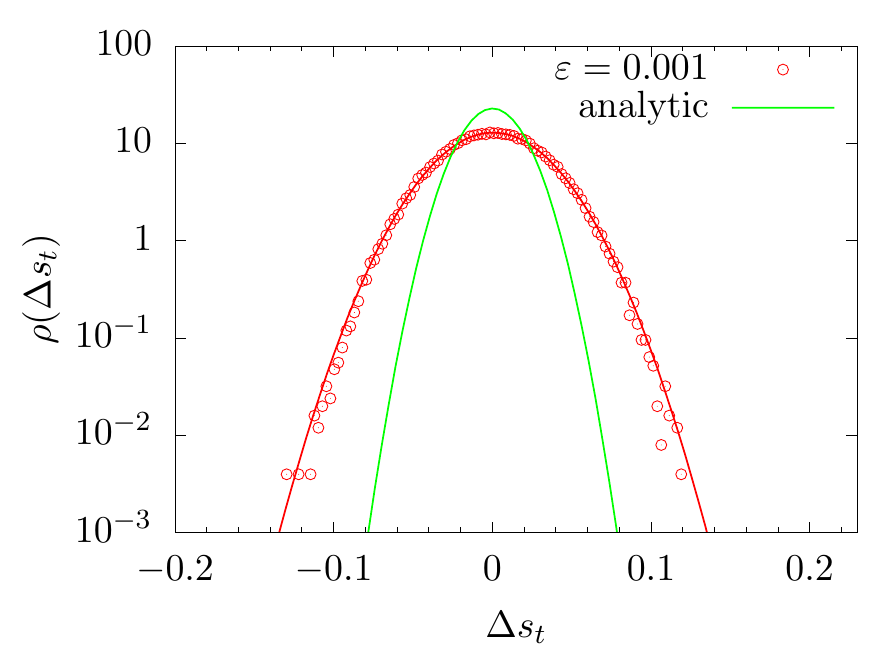}
\caption{(Color online) Data points denote simulated probability distribution of EP at dimensionless driving rate $\zeta=10^{-3}$ around a field $mH/\kb T=2$, integrated over a period of $1.28\t$. 
The (red) solid line through data denotes a fit to Eq.(\ref{rho-gauss}) giving $\la \D s_t\ra=4.8 \times 10^{-4}\kb$. 
The (green) line labeled analytic plots the function in Eq.(\ref{rho-gauss})  with $\la \D s_t\ra=1.5 \times 10^{-4}\kb$ obtained from Eq.(\ref{eq_adia}). 
}
\label{adia_st}
\end{center}
\end{figure}

In principle, the mean EP $\la \D s_t \ra$ may be obtained from Eq.(\ref{dot_St}),  using numerical methods. However, in the limit of $\zeta \ll 1$, one may use
adiabatic approximation to calculate $\la \D s_t \ra$ over a time $t$.  
In presence of a uniaxial field, Eq.(\ref{heat_1}) gives  $\dot q = -m H_\th \dot\th$. Within mean field approximation 
$\la \dot q \ra \approx -m \la H_\th \ra \la \dot \th \ra = -m^2 H^2 h' \la \sin\th \ra^2$. 
At equilibrium, the magnetization along external field $m_z/m = \la \cos \th \ra$ is given by the Langevin function,
\beq
 \la \cos \th \ra_{\rm eq} \equiv \cos \th_{\rm eq} = \le[ \coth \e -\f{1}{\e}\ri],
 \label{langefn}
 \eeq
where, $\e = m H/\kb T$.
Within adiabatic approximation, replacing $H$ by  $H(1+\zeta\, t/\t)$, one finds
$\cos \th_{\rm ad} = \cos \th_{\rm eq}  + \zeta\, g(t)$, where
$$
g(t) = \f{t}{\t}\le[ \f{1}{\e} - \f{\e}{\sinh^2 \e}\ri].
$$
Thus the mean dissipated heat is
\begin{align}
&\la \dot q \ra =  -m^2 H^2 h' (1-\cos^2 \th_{\rm ad})\nn\\
&= \zeta \, 2m^2 H^2 h' \cos \th_{\rm eq}\, g(t), \nn
\end{align}
keeping up to leading order in $\zeta$. In deriving the above relation we used the fact that $\la \dot q \ra=0$ in equilibrium. This leads to the following expressoin for
the mean total EP,  
\begin{align}
&T\la \D s_t \ra = \int_0^t dt \la \dot q \ra = \zeta (m H)^2 h'  \cos \th_{\rm eq} \f{t^2}{\t} \le[ \f{1}{\e} - \f{\e}{\sinh^2 \e}\ri],\nn\\
&{\rm i.e.,~} \f{\la \D s_t \ra}{\kb}= \zeta  \f{t^2}{\t} m H h' \le[ 1 - \f{\e^2}{\sinh^2 \e}\ri] \le[ \coth \e -\f{1}{\e}\ri].
\label{eq_adia}
\end{align}

We expect this relation to capture $\la \D s_t \ra$ and thus the probability distribution $\r(\D s_t)$ in the limit of slow driving, with the system
remaining close to equilibrium. To investigate this regime numerically, we need to estimate for which driving rates the system really remains close to equilibrium.
To this end, we obtain hysteresis curves for magnetic fields taken around a cycle by first linearly 
increasing them from $H_i$ to $H_\ell$ with a rate $\a$, and then reducing the field back to $H_i$. The time-period $t_p$ for this cyclic variation controls
the dimensionless rate $\zeta=\t_0/t_p$. In Fig.\ref{hysteresis}, we plot the average magnetization $m_z/m = \la \cos \th \ra$ for different values of $t_p$ indicated in the legend
of the figure. We use $H_i = -10\, \kb T/m$ and $H_\ell = 10\, \kb T/m$.
Note that with increasing $t_p$ the area under the curve of the hysteresis loop, a measure of energy dissipation, reduces. Finally, for
$t_p = 5\times 10^4\,\t$, the hysteresis loop collapses onto the equilibrium magnetization $ \cos \th_{\rm eq}$ given by the Langevin function in Eq.(\ref{langefn}).     
The corresponding rate is $\zeta \approx 10^{-4}$. The hysteresis loops indicate that the regime of validity for adiabatic approximation is $10^{-4} \lesssim \zeta \lesssim 10^{-3}$. 
The slowest driving rate in Fig.\ref{rho_st} is $\zeta \sim 10^{-1}$, two orders of magnitude faster than the possible regime of validity of 
the adiabatic approximation.

In Fig.\ref{adia_st} we show $\r(\D s_t)$ for 
a linear driving with dimensionless rate $\zeta=10^{-3}$ around $H=2\, \kb T/m$ over a time scale of $1.28\t$. 
Note that the simulated probability distribution predicts $\la \D s_t\ra=4.8 \times 10^{-4}\kb$ which is of the same order of magnitude 
of $\la \D s_t\ra=1.5 \times 10^{-4}\kb$ obtained from the analytic expression Eq.(\ref{eq_adia}) obtained within adiabatic approximation. The analytic estimate 
for the probability distribution of EP fails to exactly capture simulation result, as the rate of change of magnetic field is still {\em fast} with respect to the regime in which 
adiabatic approximation is strictly valid. It should be noted that, already the driving rate is very slow leading to extremely small amount of average EP  $\sim 10^{-4}\, \kb/\t$.
By taking smaller value of rate $\zeta$ one gets a better comparison, but mean EP becomes extremely small.

}

\section{Particle diffusing on the surface of a sphere}
\label{sph_diffusion}
The Langevin equation describing diffusive motion of particles on a spherical surface under external torque ${\bf N} = N_\phi \hat \phi$ acting in the azimuthal direction is
\bea
\dot \th = \mu h_\th - \mu (\sin\th)^{-1} (N_\phi \sin \th + h_\phi) \nn\\
\sin \th\, \dot \phi = \mu h_\th + \mu (\sin\th)^{-1}  (N_\phi \sin \th + h_\phi),
\eea
where $\mu$ is the mobility of the particle. In this over damped dynamics kinetic energy is absent, and for non-interacting particles potential energy is also zero.
The stochastic energy balance allows one to express the rate of work done by the external torque $\dot W$ in terms of the dissipated heat $\dot q$ as 
\beq
\dot W = N_\phi \sin\th\, \dot \phi = - \dot q.
\label{1st_law_2}
\eeq

In the absence of torque $N_\phi=0$ the equations describe simple diffusion on the surface of a unit sphere. The corresponding FP equation 
$\p_t P = k' \D_\o P$ where the Laplace-Beltrami operator $\D_\o = (1/\sin\th) \p_\th (\sin\th \p_th) + (1/\sin^2\th) \p_\phi^2$ and $k' = \mu \kb T$. The solution of this equation
can be easily found using spherical harmonics obeying $\D_\o Y_{l m }(\o) = -l(l+1) Y_{lm}(\o)$. Expanding $P(\o,t)$ in the spherical harmonic basis, one finds 
$P(\o,t) = \sum_{lm} a_{lm} e^{-l(l+1)k' t} Y_{lm}(\o)$. The constant $a_{lm}$ depends on the initial condition. If one choses a delta-function distribution $\d(\o - \o')$ as initial condition, one finally
obtains $P(\o,t) = \sum_{lm} Y^\ast_{lm}(\o') e^{-l(l+1)k' t} Y_{lm}(\o)$.

In the presence of ${\bf N} = N_\phi \hat \phi$ the FP equation is given by $\p_t P = - \nabla_\o . {\bf J}_\o$ with ${\bf J}_\o = \hat \th J_\th + \hat \phi J_\phi$
where
\bea
J_\th &=& -\mu N_\phi P -k' \p_\th P \nn\\
J_\phi &=& \mu N_\phi P -k' (\sin \th)^{-1} \p_\phi P.
\eea
The natural variables denoting a microstate for a diffusing particle is the angular position coordinates $(\th,\phi)$. Under time reversal 
they transform as even variables. 
Thus, unlike in the case of macrospins, the complete expression of currents $J_\th$ and $J_\phi$ are 
dissipative currents.  This leads to a new form of EP that has a non-Clausius {excess} EP. 

\subsection{Entropy production using Fokker-Planck equation}
The rate of change of system entropy 
\bea
\f{\dot s}{\kb} &=& -\f{\p_t P}{P} - \f{\p_\th P}{P} \dot \th - \f{\p_\phi P}{P} \dot \phi \nn\\
&=& -\f{\p_t P}{P} + \f{J_\th \dot \th + J_\phi \sin\th\, \dot\phi}{k' P} - \f{\dot s_r}{\kb}, 
\label{sdot_2}
\eea
where, 
\bea
\dot s_r = \f{1}{T} N_\phi \le[ \sin\th\, \dot\phi -  \dot \th \ri] =  -\f{\dot q}{T} - \f{N_\phi \dot \th}{T}.
\label{srdot_2}
\eea 
In the last step, we used Eq.(\ref{1st_law_2}) to express the reservoir EP in terms of dissipated heat $\dot q$. 
The amount $- N_\phi \dot \th/T$ is the measure of {\em excess} EP, an EP excess to the Clausius measure of $-\dot q/T$.
 
From Equation (\ref{sdot_2}), using the two step averaging as in Sec.\ref{ep_fp} it is straightforward to show that the average total entropy production
\bea
\dot S_t = \la \dot s \ra + \la \dot s_r \ra = \kb \int d\o \f{J_\th^2 + J_\phi^2}{k'P} \geq 0
\eea
in accordance with the second law of thermodynamics. 

\subsection{Fluctuation theorems}
The probability of time forward trajectories denoted by $X = [\th(t),\phi(t),{\bf N}(t)]$  remains same as ${\cal P}_+$ shown in Sec.\ref{ft_1}. However, given that $\th,\,\phi$ are even variables under 
time reversal, probability of time-reversed trajectory ${\cal P}_-$ changes. The external torque ${\bf N}(t)=\hat\phi N_\phi(t)$ is a control parameter which traces back under time-reversal without changing sign, 
$X^\dagger = [\th(\t_0 - t), \phi(\t_0 -t), {\bf N}(\t_0 -t)]$. The probability of conjugate trajectory   
${\cal P}_- = \prod_{i=1}^N p_i^-$, where 
$p_i^- =    \J_i^- \la \d(\dot \th_i +\Theta_i(\t_0 -t)\,)\, \d(\dot \phi_i + \Phi_i(\t_0 -t) \ra $, 
where $ \J_i^-$ denotes the relevant Jacobian.  As $\J_i^-=\J_i$, Jacobians drop out of the ratio $p_i^+ / p_i^-$.
After some algebra, it is possible to show that 
the ratio of two probabilities of forward and reverse paths $\f{{\cal P}_+}{{\cal P}_- } = \exp(\D s_r/\kb)$, where 
\bea
\f{\D s_r}{\kb} =  \f{1}{\kb T} \int_0^{\t_0} dt  N_\phi \le[ \sin\th \, \dot \phi - \dot \th \ri],
\label{dsr2}                                        
\eea
i.e., $\D s_r$ leads to the rate of EP $\dot s_r$  [Eq.(\ref{srdot_2})] derived from FP equation.
As before, it is straight forward to show that the total entropy production $\D s_t = \D s + \D s_r$ obeys the IFT $\la e^{-\D s_t/\kb} \ra =1$ and the DFT $\r(-\D s_t) = e^{-\D s_t/\kb} \r(\D s_t)$. 

\section{Discussion}
Note that essentially the same Langevin and FP equations 
describe the dynamics of both a macrospin under external magnetic field, and diffusion of a particle on a unit sphere under external torque. 
The stochastic equation of motion directly leads to stochastic energy balance, defining the expression of dissipated heat. 
However, the variables defining microstates and their symmetry  under time-reversal (odd or even) is different in the two cases. This leads to different 
expressions for irreversible currents~\cite{Spinney2012}, and as a result different expressions for EP in the environment. 
While for macrospins reservoir EP is given entirely by the Clausius expression, for particle diffusing on unit sphere one obtains an excess EP in addition to Clausius like term. 
One arrives at the same conclusion by using probability ratio of forward and time-reversed trajectories. 
It is interesting to note that, even if the external torque is time-independent, EP for particles diffusing on unit sphere can be non-zero, with probability distributions $\r(\D s_t)$ obeying the DFT, unlike macrospins in which EP remains zero if the external field $\H$ is time-independent. We showed that the total stochastic EP obeys fluctuation theorems. In particular, we analyzed stochastic dynamics
of macrospins numerically, to obtain probability distributions of EP which becomes broad and Gaussian for slow rate of change of the external magnetic field. We obtained analytic expression of the 
distribution in the adiabatic limit and presented its comparison with numerical results.

\acknowledgments
SB and DC thank Sashideep Gutti for useful discussions. AMJ thanks DST, India for financial support.

\bibliographystyle{prsty}


\end{document}